\documentclass[aps,prd,twocolumn,english,notitlepage,nofootinbib,superscriptaddress]{revtex4-1}
\usepackage[T1]{fontenc}
\usepackage[latin9]{inputenc}
\usepackage[usenames,dvipsnames]{color}  
\definecolor{darkred}{rgb}{0.6,0,0}
\definecolor{brown}{rgb}{0.59, 0.29, 0.0}
\definecolor{xunjie}{rgb}{0.059,0.52,0.75}
\usepackage{amsmath}
\usepackage{graphicx}
\usepackage[colorlinks=true,citecolor=darkred,urlcolor=darkred, pdfborder={0 0 0}]{hyperref}
\usepackage{ulem}
 
\newcommand {\ignore}[1]{}
\def\321{$\mathrm{SU(3) \otimes SU(2) \otimes U(1)}$ }

\def\lsim{\raise0.3ex\hbox{$\;<$\kern-0.75em\raise-1.1ex\hbox{$\sim\;$}}}

\def\clfv{charged lepton flavour violation }
\def\clfvg{charged lepton flavour violating }

\def\SM{$\mathrm{SU(3)_c \otimes SU(2)_L \otimes U(1)_Y}$ }

\def\gsim{\raise0.3ex\hbox{$\;>$\kern-0.75em\raise-1.1ex\hbox{$\sim\;$}}}
\def\lsim{\raise0.3ex\hbox{$\;<$\kern-0.75em\raise-1.1ex\hbox{$\sim\;$}}}

\def\3211{$\mathrm{SU(3) \otimes SU(2)_L \otimes U(1)_R \otimes U(1)_{B-L}}$ }
\def\321{$\mathrm{SU(3) \otimes SU(2) \otimes U(1)}$ }
\def\422{$\mathrm{SU(4) \otimes SU(2) \otimes SU(2)_R}$ }

\definecolor{linkcolor}{rgb}{0,0,0.5}

\makeatletter

\usepackage{slashed}

\makeatother

\usepackage{babel}
\begin{document}

\title{\color{BrickRed}
   High-energy colliders as a probe of neutrino properties }

\author{Sanjoy Mandal}\email{smandal@kias.re.kr}
\affiliation{Korea Institute for Advanced Study, Seoul 02455, Korea}

\author{O. G. Miranda}\email{omr@fis.cinvestav.mx}
\affiliation{Departamento de F\'{\i}sica, Centro de   Investigaci{\'o}n y de Estudios Avanzados del IPN\\ Apartado Postal   14-740 07000 Ciudad de M\'exico, Mexico}

\author{G. Sanchez Garcia}\email{gsanchez@fis.cinvestav.mx}
\affiliation{Departamento de F\'{\i}sica, Centro de   Investigaci{\'o}n y de Estudios Avanzados del IPN\\ Apartado Postal   14-740 07000 Ciudad de M\'exico, Mexico}

\author{J. W. F. Valle}\email{valle@ific.uv.es}
\affiliation{AHEP Group, Institut de F\'{i}sica Corpuscular --   C.S.I.C./Universitat de Val\`{e}ncia, Parc Cientific de Paterna.\\   C/Catedratico Jos\'e Beltr\'an, 2 E-46980 Paterna (Val\`{e}ncia) - SPAIN}

\author{Xun-Jie Xu}\email{xuxj@ihep.ac.cn}
\affiliation{Institute of High Energy Physics, Chinese Academy of Sciences, Beijing 100049, China}.

\date{\today}
\begin{abstract}
The mediators of neutrino mass generation can provide a probe of
neutrino properties at the next round of high-energy hadron~(FCC-hh)
and lepton colliders~(FCC-ee/ILC/CEPC/CLIC).
We show how the decays of the Higgs triplet scalars mediating the
simplest seesaw mechanism can shed light on the neutrino mass scale
and mass-ordering, as well as the atmospheric octant.
Four-lepton signatures at the high-energy frontier may provide the
discovery-site for charged lepton flavour non-conservation in nature,
rather than low-energy intensity frontier experiments.
    
\end{abstract}
\maketitle

\section{Introduction}
\label{sec:introduction}

Solar and atmospheric neutrino studies provided the discovery site for neutrino oscillations~\cite{Kajita:2016cak,McDonald:2016ixn}.
These experiments were followed by reactor~\cite{DayaBay:2012fng} and accelerator-based~\cite{K2K:2006yov} studies that have confirmed the oscillation phenomenon
and also substantially improved parameter determination.
The discovery of neutrino oscillations has brought neutrino physics to the center of particle physics, giving the first clear evidence for new physics.
Their existence also suggests the possibility of charged lepton flavour violating effects including rare processes, such as $\mu\to e\gamma$ decays~\cite{MEG:2013oxv}.

Although current neutrino data are well-described by the \textit{three-neutrino paradigm}, there are still loose ends to sort out,
such as the neutrino mass-ordering, the atmospheric octant and the precise value of the CP phase~\cite{deSalas:2020pgw,10.5281/zenodo.4726908}. 
 Back in the LEP days it was suggested that the mediators of neutrino mass generation could be produced at collider experiments~\cite{Dittmar:1989yg} in such a way that
 high-energy studies could be used to probe neutrino oscillation parameters, such as the atmospheric angle.
  This is a characteristic feature, e.g., of models where supersymmetry is the origin of neutrino mass~\cite{Romao:1991ex,Hirsch:2000ef,Diaz:2003as,Hirsch:2004he},
  which allow for such independent probes of neutrino mixing~\cite{deCampos:2007bn,DeCampos:2010yu,deCampos:2012pf}. 

 In this letter we propose the use of high-energy frontier hadron~(FCC-hh~\cite{FCC:2018vvp}) and lepton colliders~(FCC-ee~\cite{FCC:2018evy}/ILC~\cite{Barklow:2015tja}/
  CLIC~\cite{CLICdp:2018cto}, CEPC~\cite{CEPCStudyGroup:2018rmc}),
  as an independent probe of neutrino properties, capable of shedding light on the neutrino mass scale and the neutrino mass-ordering through the rates for four lepton final-state events, as well as the atmospheric octant can be probed through the triplet Higgs decay pattern.
  Moreover, such high-energy  experiments can also provide the first evidence for charged lepton flavour violation in nature.

\section{Origin of neutrino mass}
\label{sec:origin-neutrino-mass}
Despite efforts to underpin the ultimate mechanism responsible for neutrino mass generation, the challenge remains wide open. An attractive possibility is provided by the seesaw mechanism.
Its most general formulation employs the Standard Model (SM) picture, i.e. the \SM gauge group~\cite{Schechter:1980gr,Schechter:1981cv}.
This leads to small active neutrino masses induced through the exchange of heavy lepton or scalar mediators. 
 The simplest seesaw mechanism is the type II seesaw with explicit breaking of lepton number, where neutrino masses are
 mediated by a triplet scalar, 
\begin{equation}
\Delta = \frac{1}{\sqrt{2}}\begin{pmatrix}
 \Delta^{+}& \sqrt{2}\Delta^{++}\\ 
 v_{\Delta} + h_{\Delta} + i\eta_{\Delta}& - ~\Delta^{+}
\end{pmatrix},~
\end{equation}
so that only one complex symmetric Yukawa matrix $Y_{\Delta\alpha \beta}$ describes the full flavour structure of the lepton sector~\cite{Schechter:1980gr,Schechter:1981cv}
through the Yukawa Lagrangian term
\begin{equation}
{\cal L}_{{\rm type\ II}}=\left[iY_{\Delta\alpha\beta}L_{\alpha}^{T}C^{-1} \tau_2\Delta L_{\beta}+\text{h.c.}\right]
\label{app:x-3}
\end{equation}
where $L_{\alpha}$ are the lepton doublets, $C$ is the charge conjugation operator. The scalar potential $V(\Phi,\Delta)$ is given as, 
\begin{align}
& V(\Phi,\Delta) =  -m_{\Phi}^{2}\Phi^{\dagger}\Phi + \frac{\lambda}{4}(\Phi^{\dagger}\Phi)^{2}  +  \tilde{M}_{\Delta}^{2}{\rm Tr}\left[\Delta^{\dagger}\Delta\right] \nonumber \\
&+\lambda_{2}\left[{\rm Tr}\Delta^{\dagger}\Delta\right]^{2} +\lambda_{3}{\rm Tr}\left[\Delta^{\dagger}\Delta\right]^{2} + \left[\mu \Phi^{T}i\sigma_{2}\Delta^{\dagger}\Phi+\text{h.c.}\right] \nonumber \\
&+\lambda_{1}(\Phi^{\dagger}\Phi){\rm Tr}\left[\Delta^{\dagger}\Delta\right]+\lambda_{4}\Phi^{\dagger}\Delta\Delta^{\dagger}\Phi.\label{eq:potential}
\end{align} 
where $\Phi$ is the SM Higgs doublet. Its minimization generates a non-zero vacuum expectation value (VEV) for the neutral component of the
triplet. Within the simplest approximation~($\tilde{M}_{\Delta}\gg v_\Phi$) the small induced triplet VEV $v_{\Delta}$ is given as  
\begin{equation}
v_{\Delta} \approx \frac{\mu v_{\Phi}^{2}}{\sqrt{2}\tilde{M}_{\Delta}^{2}}, \label{V-triplet-approx}
\end{equation}
where $v_\Phi$ is the SM Higgs VEV. Eq.~(\ref{V-triplet-approx}) shows how the smallness of $v_{\Delta}$ requires either by a small $\mu$, or a large value for $\tilde{M}_\Delta$ characterizing the triplet scalar mass~\cite{Inprep}.
Note that in a more complete setup the parameter $\mu$ can be given a full dynamical interpretation~\cite{Schechter:1981cv,Bonilla:2015jdf}.
Following t'Hooft's naturalness argument, the small $\mu$ parameter sources lepton number violation,
so that in the limit $\mu \to 0$, lepton number symmetry is recovered. 
After electroweak symmetry breaking one obtains small neutrino masses
\begin{equation}
  m_{\alpha\beta}^{\nu}\equiv (Y_{\Delta})_{\alpha\beta}\frac{v_{\Delta}}{\sqrt{2}}. \label{mnu-triplet-approx}
\end{equation}
For the simplest CP-conserving case the neutrino oscillation parameters will be determined by six elements of the real symmetric Yukawa matrix $Y_{\Delta\alpha \beta}$.
Four of these are well measured~\cite{deSalas:2020pgw,10.5281/zenodo.4726908}, the remaining ones being the absolute neutrino mass and the atmospheric octant. 

There are seven physical Higgs fields with definite masses namely,
a doubly-charged~($H^{\pm\pm}$) and a singly-charged~($H^{\pm}$) scalar boson,
plus two massive CP-even~($h,H$) and one massive CP-odd Higgs~($A$).
The new scalars present in such simplest seesaw scheme may hold the key to electroweak vacuum stability and perturbative unitarity~\cite{Bonilla:2015eha,Bonilla:2015jdf}.
The associated theoretical consistency restrictions, as well as the constraints from electroweak precision data and other experiments are discussed extensively in~\cite{Inprep}. 


\section{simplest seesaw at colliders}
\label{sec:compl-betw-clfv}

There have been several experimental searches for signatures associated to the doubly-charged Higgs at high energy colliders~\cite{ATLAS:2017xqs,CMS:2017pet,ATLAS:2018ceg}.
Theoretical aspects of the type-II seesaw and triplet scalar models at different colliders have been discussed in
\cite{Arhrib:2011uy,Mitra:2016wpr,Antusch:2018svb,Du:2018eaw,Padhan:2019jlc,Primulando:2019evb,Ashanujjaman:2021txz} and reviewed in~\cite{Cai:2017mow}.
Depending on the magnitude of the triplet VEV $v_\Delta$, the doubly-charged Higgs mainly decays to same-sign dileptons~($v_\Delta\leq 10^{-4}$ GeV) or gauge bosons~($v_\Delta > 10^{-4}$ GeV).
Studying these leptonic or gauge boson decay modes at the LHC, one can constrain doubly charged Higgs properties.
For small triplet VEV $v_\Delta\leq 10^{-4}$~GeV, one obtains $m_{H^{\pm\pm}}>870$ GeV \cite{CMS:2017pet,ATLAS:2017xqs}, whereas for $v_\Delta>10^{-4}$ GeV,
the constraint is rather loose, $m_{H^{\pm\pm}}>220$ GeV \cite{ATLAS:2018ceg}.

\begin{figure}[t]
\begin{center}
\includegraphics[height=3.5cm,width=0.5\textwidth]{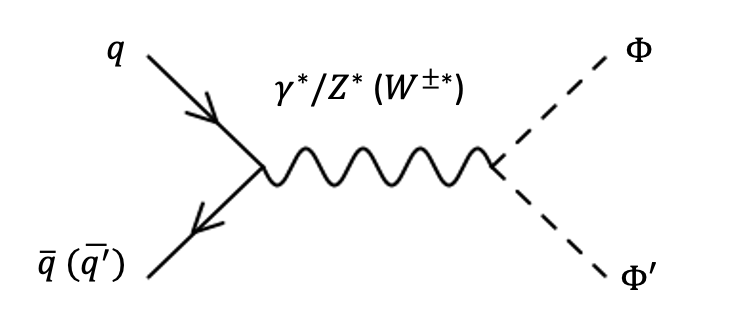}
\end{center}
\caption{Feynman diagrams for Drell-Yan production of triplet Higgs scalar bosons $\Phi,\Phi'\in\{H^0,A,H^\pm,H^{\pm\pm}\}$ within the type-II seesaw mechanism.}
    \label{fig:feyn-diagH}
\end{figure}

\begin{figure}[t]
\begin{center}
\includegraphics[width=0.4\textwidth]{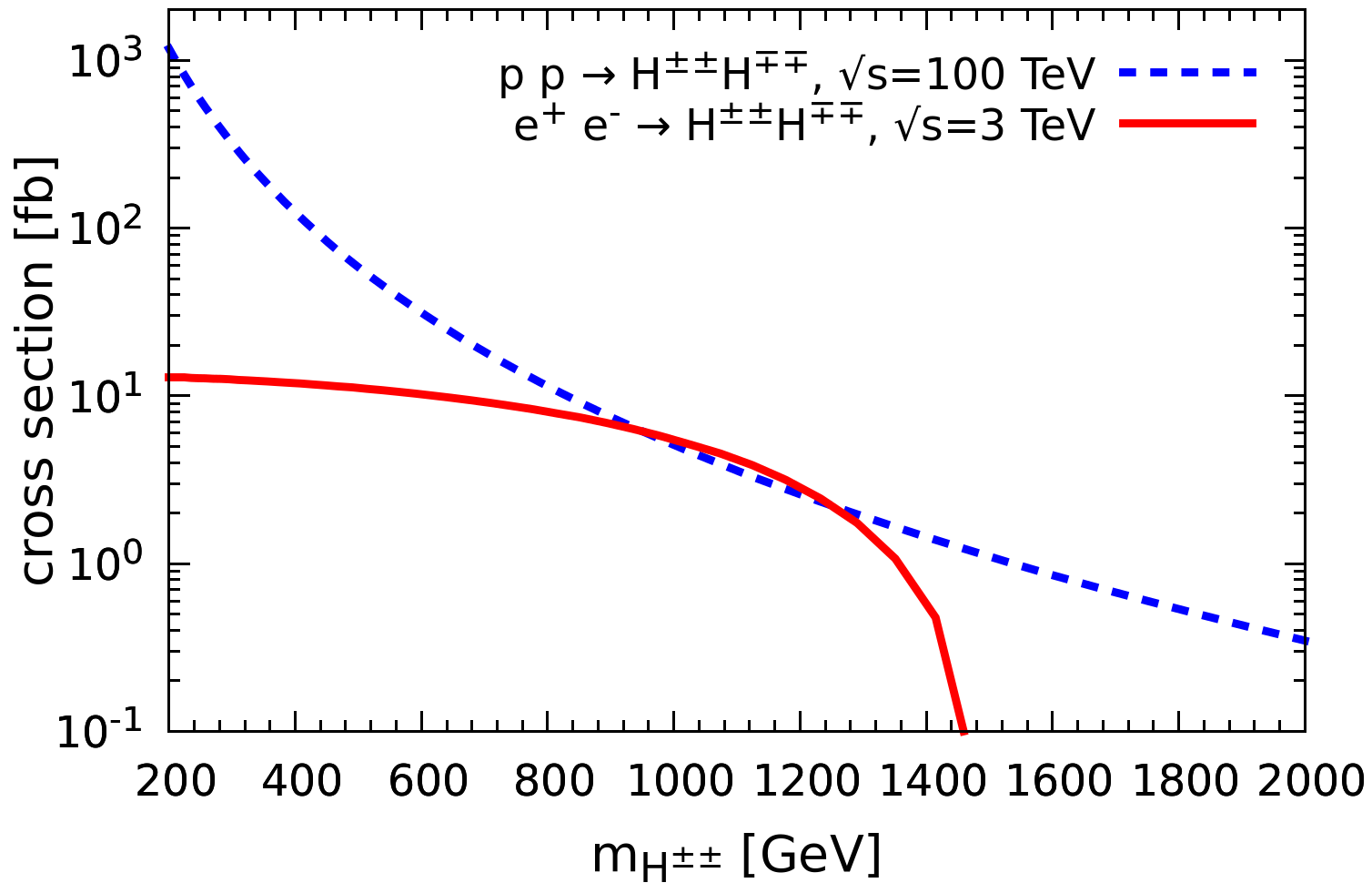}
\end{center}
\caption{
  $H^{\pm\pm}H^{\mp\mp}$ production cross section at a 3 TeV $e^+e^-$ collider~(red solid) and a 100 TeV $pp$ collider~(blue-dashed).}
    \label{fig:Xsection}
  \end{figure}

A key feature of the simplest seesaw mechanism is the presence of a doubly charged Higgs scalar $H^{\pm\pm}$ which can be produced via the Drell-Yan process $pp,\,e^+e^- \to \gamma^*/Z^* \to H^{\pm\pm}H^{\mp\mp}$, see Fig.~\ref{fig:feyn-diagH}.
%
In Fig~\ref{fig:Xsection}, we show this cross section at the proposed FCC-hh~($\sqrt{s}=100$~TeV) $pp$-collider and a $\sqrt{s}=3$~TeV $e^{+}e^{-}$ ILC/CEPC/CLIC/FCC-ee collider.
  One sees that multi-TeV Higgs masses can be explored at such large center-of-mass energies
  and that for $m_{H^{\pm\pm}}\sim 1$~TeV, the production cross section is similar both for the 100 TeV $pp$ and 3 TeV $e^+e^-$ colliders.
  We will choose $m_{H^{\pm\pm}}=1$~TeV as the benchmark for our study.

  Once produced, the neutrino mass mediator $H^{\pm\pm}$ can decay to two same-sign charged leptons ($l^{\pm}l^{\pm}$). 
  The $H^{\pm\pm}$ decay branching ratio to charged leptons depends on the Yukawa coupling $Y_{\Delta \alpha \beta}$ and hence on the triplet VEV $v_{\Delta}$. 
  For $v_{\Delta}<10^{-4}$ GeV, $H^{\pm\pm}\to l^{\pm}l^{\pm}$ is the dominant decay mode.
The flavor structure of $Y_{\Delta \alpha \beta}$ is of crucial
importance here, as it determines both the neutrino properties as well
as the branching ratio of $H^{\pm\pm}$ to charged leptons of different
flavors at colliders.

The doubly-charged Higgs boson decays to leptonic final states are determined by the Yukawa coupling $Y_\Delta$. The decay widths are given by
\begin{align}
\Gamma (H^{\pm \pm} \to l^{\pm}_i l^{\pm} _j)=\frac{m_H^{\pm \pm} } {(1+\delta_{ij}) 16 \pi}  |Y_{\Delta}^{ij}|^2,
\end{align}
with $Y_\Delta=\frac{\sqrt{2}}{v_\Delta} U\textup{diag}\left\{ m_{\nu_1}, m_{\nu_2}, m_{\nu_3}\right\} U^T$, where $U$ is the lepton mixing matrix measured in oscillation experiments. 
The patterns of various leptonic channels will follow the profile of $Y_\Delta^{ij}$. 
Current measurements of neutrino oscillation parameters~\cite{deSalas:2020pgw,10.5281/zenodo.4726908} restrict the diagonal and off-diagonal entries of the Yukawa coupling matrix $Y_\Delta$.
Depending on the ordering of the light-neutrino mass spectrum we obtain the following decay branching ratio patterns: 
\begin{align}
&\text{BR}(H^{++}\to \mu\mu), \text{BR}(H^{++}\to \tau\tau)\gg \text{BR}(H^{++}\to ee) \text{ {\bf NO}},\nonumber\\
& \text{BR}(H^{++}\to ee)\gg \text{BR}(H^{++}\to \mu\mu), \text{BR}(H^{++}\to \tau\tau) \text{ {\bf IO}},\nonumber
\end{align}
suggesting that, depending on the ordering of the light neutrino masses, $H^{++}$ can mainly decay to $\mu\mu$, $\tau\tau$~(for {\bf NO}) or $ee$~({\bf IO}). 
Hence, it may be possible to probe the neutrino mass ordering normal ordering~({\bf NO}) or inverted ordering~({\bf IO}) just by looking into the decays of $H^{++}$ to same-flavour leptonic final states.
\begin{figure}[t]
\begin{center}
\includegraphics[width=0.45\textwidth]{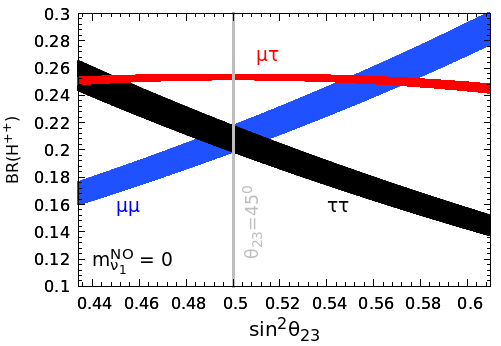}
\includegraphics[width=0.45\textwidth]{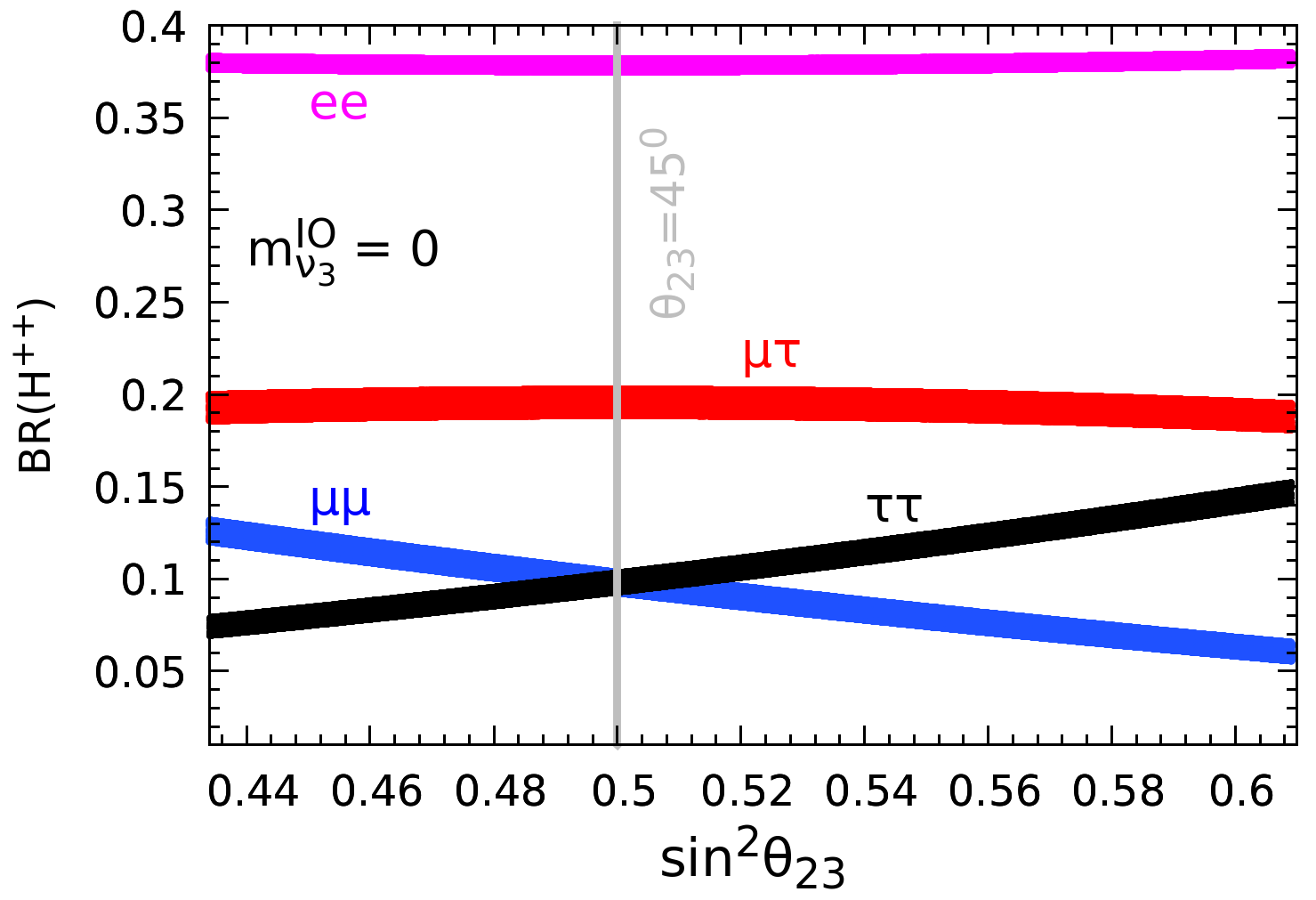}
\end{center}
\caption{ 
  $H^{\pm\pm}$ decay branching ratios versus the atmospheric angle within its allowed $3\sigma$ range, for both {\bf NO}~(top panel) and {\bf IO}~(bottom panel), repectively.
  The lightest neutrino mass is fixed to zero, with triplet vev $v_\Delta<10^{-4}$~GeV, so that the dilepton $H^{\pm\pm}$ decay channel is dominant.
  The Dirac phase $\delta_{\text{CP}}$ is varied in the range $[-\pi:\pi]$ and other oscillation parameters are fixed to their best fit values~\cite{deSalas:2020pgw,10.5281/zenodo.4726908}. }
    \label{fig:Atmos}
  \end{figure}
  In Fig.~\ref{fig:Atmos}, we show the leptonic branching ratios~(only those which are appreciable) both for {\bf NO}~(top panel) and {\bf IO}~(bottom panel).
 The results are shown with respect to the $3\sigma$ allowed range of the ``atmospheric'' mixing angle $\theta_{23}$. 
 Here the triplet vev is chosen as $v_\Delta<10^{-4}$~GeV, $\delta_{\text{CP}}$ is varied in the range $[-\pi:\pi]$, and other oscillation parameters are fixed to their best fit
 values~\cite{deSalas:2020pgw,10.5281/zenodo.4726908}.
 The vertical line in each panel denotes $\theta_{23}=45^\circ$. 
 One sees that for {\bf NO}, electron final-states are penalized with respect to those into muons and taus 
 i.e. one can have sizeable branching ratios for $\mu\mu$ and $\mu\tau$ final states. One the other hand, for {\bf IO}, $ee$ final states are dominant. 
\begin{figure}[t]
\begin{center}
\includegraphics[width=0.4\textwidth]{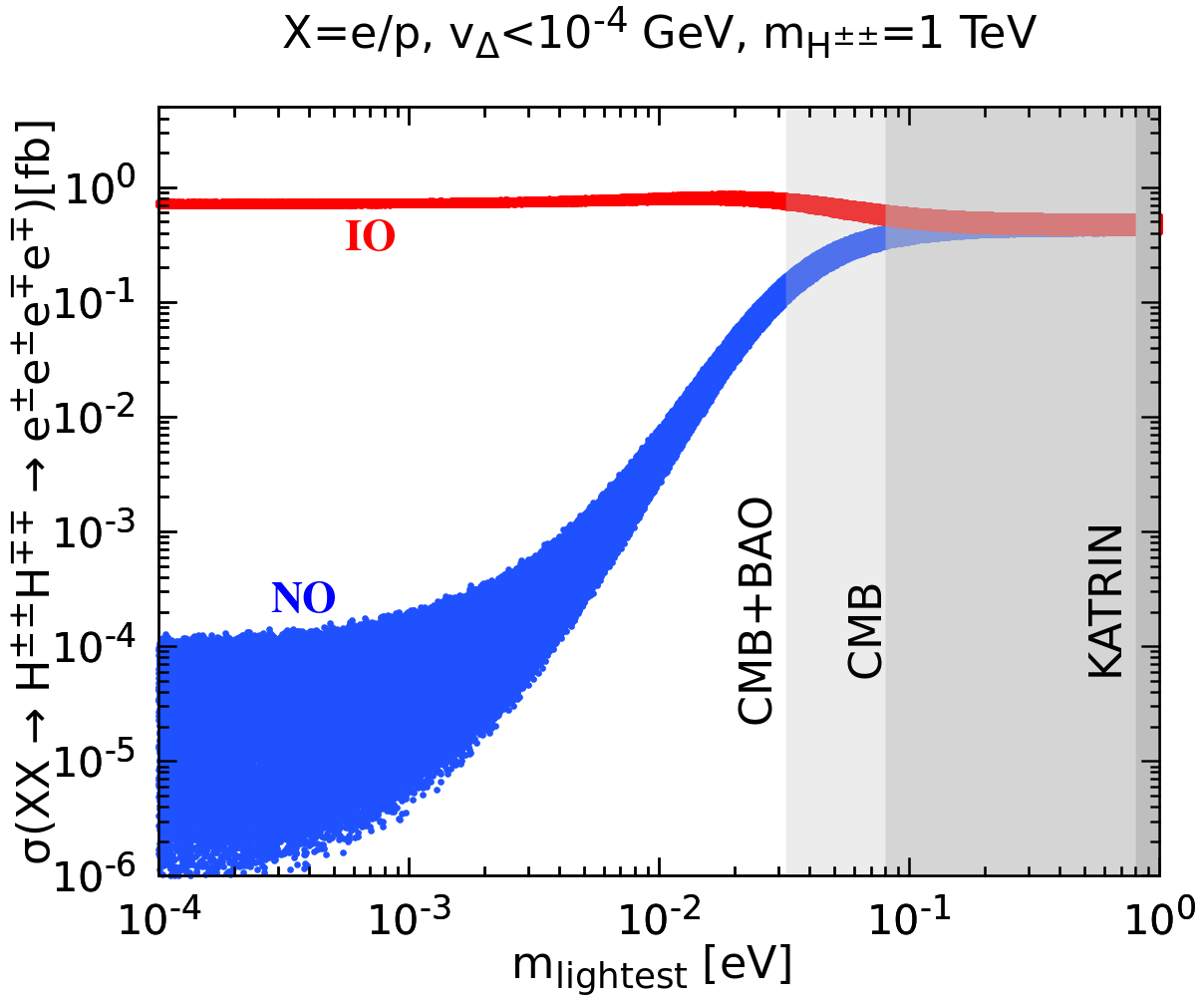}
\includegraphics[width=0.4\textwidth]{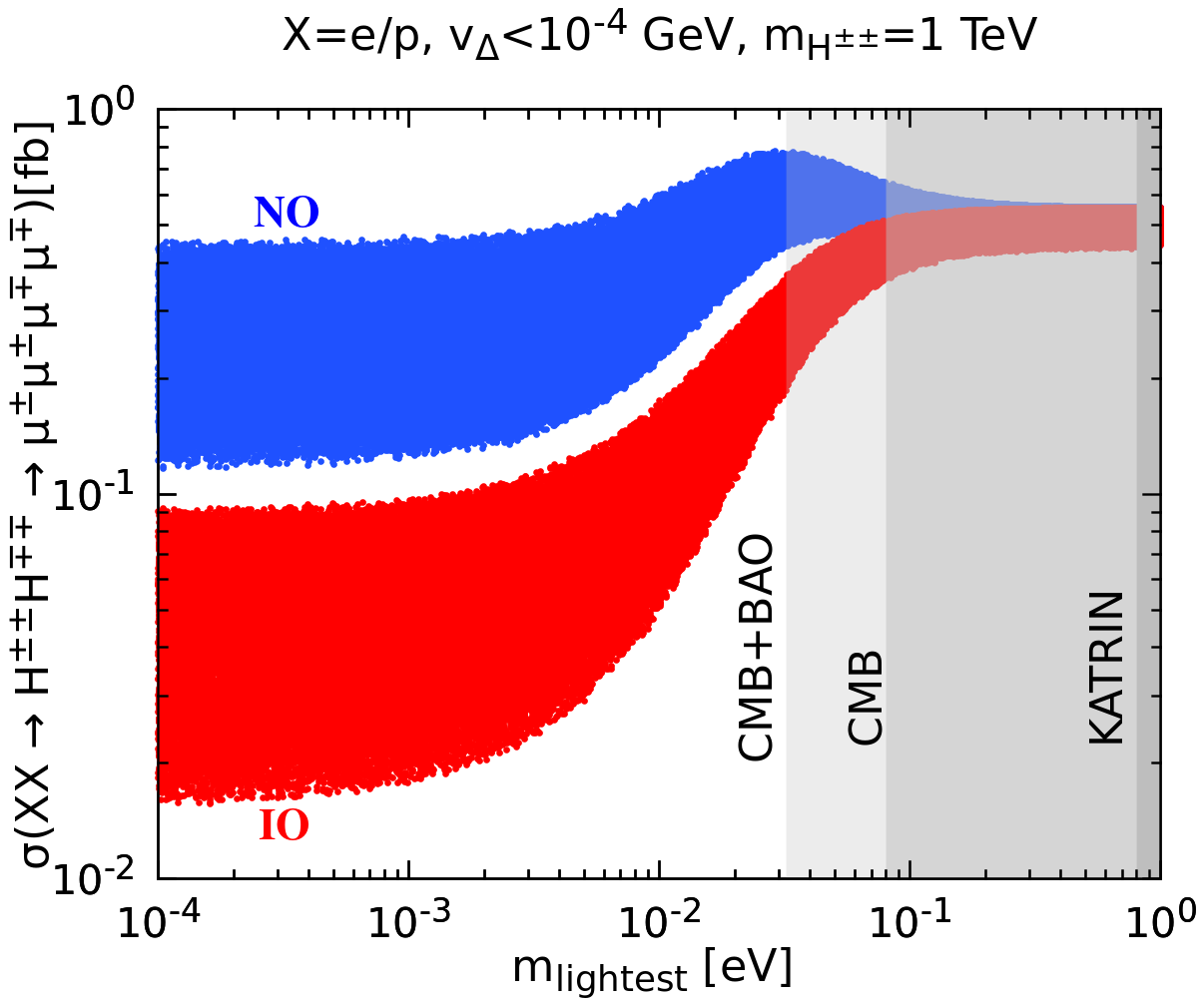}
\end{center}
\caption{ 
  Probing the neutrino mass ordering through the four-lepton cross section.
  Red points correspond to {\bf IO}, while blue points correspond to {\bf NO}.
  Top and bottom panels correspond to the tetra-electron and tetra-muon final states , respectively.
  As before, we varied the oscillation parameters in their $3\sigma$ ranges~\cite{deSalas:2020pgw,10.5281/zenodo.4726908}, fixing the
doubly-charged Higgs mass at $m_{H^{\pm\pm}}=1$~TeV, so that both the
3 TeV $e^+e^-$ and 100 TeV $pp$ colliders have similar cross
section~(5 fb).
The vertical shaded bands show the KATRIN, CMB, and CMB+BAO lightest-neutrino-mass limits.}
    \label{fig:BRLFVandCross}
  \end{figure}

We now turn to the full 4-lepton signal cross-sections, starting from the lepton-flavour conserving ones.
For large triplet VEV $v_\Delta > 10^{-4}$~GeV, the Yukawa coupling is small and $H^{\pm\pm}$ decays predominantly to $W^{\pm}W^\pm$.
However, for small $v_\Delta$, the Yukawa coupling is large and $H^{\pm\pm}$ decays predominantly to dileptons, hence the four-lepton cross section can be experimentally detectable. 
In Fig.~\ref{fig:BRLFVandCross} we illustrate the correlation between $\sigma(e^+ e^-\to H^{\pm\pm}H^{\mp\mp}\to eeee/\mu\mu\mu\mu)$ and the lightest neutrino mass both for {\bf NO} and {\bf IO} spectra.  
One sees that the~tetra-muon cross section is always larger in the case of {\bf NO},
while the cross section~into tetra-electron final states is always larger in the case of {\bf IO}.   
Note that the difference becomes larger as the lightest neutrino mass gets smaller. 
This provides a way to probe the neutrino mass-ordering through the four-lepton final states coming from doubly-charged Higgs $H^{\pm\pm}H^{\mp\mp}$ pair-production. 

\section{Lepton flavour violation }
\label{sec:lept-flav-viol}

The oscillation discovery has brought neutrino physics to the center of particle physics, giving the first clear evidence for new physics namely,
lepton flavour non-conservation in neutrino propagation.
This suggests also the possibility of charged lepton flavour violating effects including rare decays such as $\mu\to e\gamma$, so far never observed~\cite{MEG:2013oxv}.
It is straightforward to determine the corresponding expressions for the $\mu\to e\gamma$ branching ratio~\cite{Cai:2017jrq,Akeroyd:2009nu}:
\small
\begin{align}
 \text{BR}(\mu \to  e\gamma) \approx \frac{\alpha}{192\pi}\frac{\left | (Y_{\Delta}^{\dagger}Y_{\Delta})_{e\mu} \right |^{2}}{G_{F}^{2}}\left ( \frac{1}{m_{H^{+}}^{2}} + \frac{8}{m_{H^{++}}^2} \right )^{2}.
\label{eq:BRmutoegamma}
\end{align}
\normalsize
\begin{figure}[t]
\begin{center}
\includegraphics[width=0.4\textwidth]{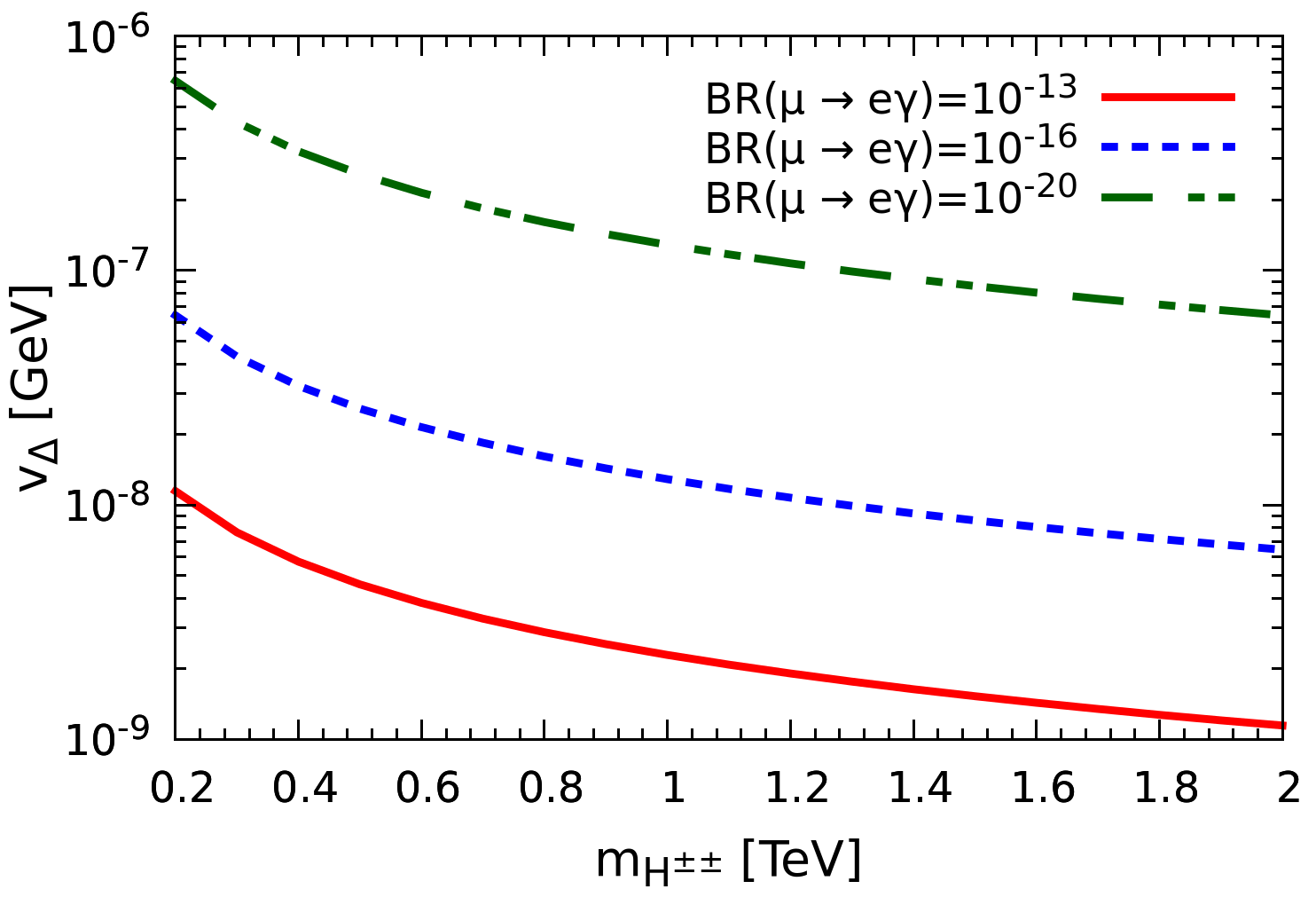}
\end{center}
\caption{Contour of BR($\mu\to e\gamma$) in the $m_{H^{\pm\pm}} - v_\Delta$ plane. The oscillation parameters are fixed to their best fit values.}
\label{fig:mutoegamma}
\end{figure}
In Fig.~\ref{fig:mutoegamma} we display some $\text{BR}(\mu\to e\gamma)$ contours in the $m_{H^{\pm\pm}}-v_\Delta$ plane, obtained for normal
ordered light neutrino spectrum and best-fit values for the neutrino oscillation parameters.  
One sees that the $\mu\to e\gamma$ branching ratio can easily exceed current senstivities~\cite{MEG:2013oxv} for small values of the triplet VEV $v_\Delta$.

The idea that charged lepton flavour (and CP) violation could be first seen at high energies was first put forward in~\cite{Bernabeu:1987gr,Rius:1989gk} and revived in~\cite{Deppisch:2013cya}.  
Here we demostrate that the type II seesaw provides the simplest realization of this idea.

Indeed we see that, from Fig.~\ref{fig:Atmos}, both for {\bf NO} and {\bf IO} one has quite large $\mu\tau$ branching fractions for $H^{++}$,
similar to those for same-flavour ($\mu\mu$) final state. On the other hand, the branching ratio to the $e\mu$, $e\tau$ final states can exceed~$\mathcal{O}(10^{-2})$.
\begin{figure}[t]
\begin{center}
\includegraphics[width=0.4\textwidth]{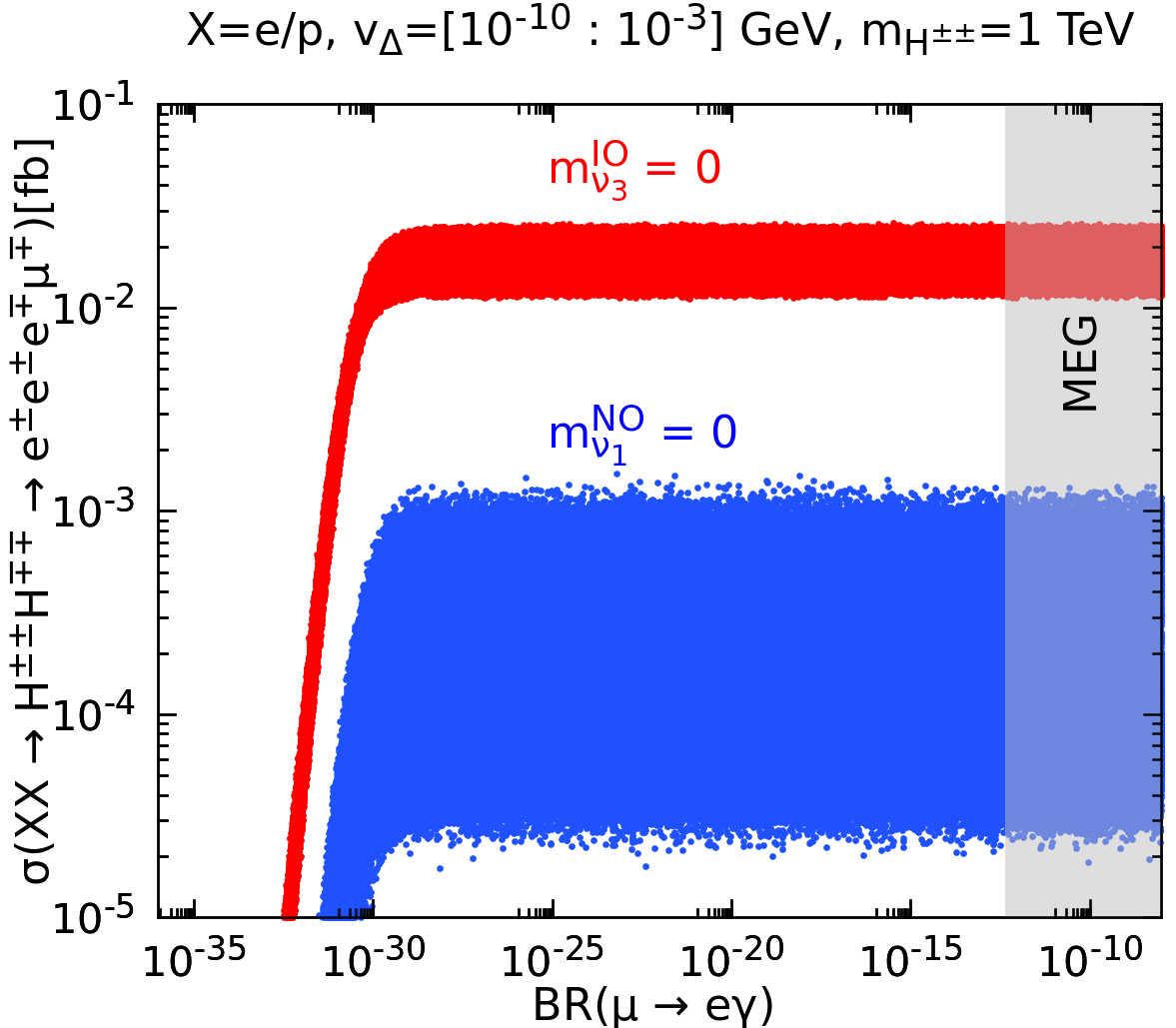}
\includegraphics[width=0.4\textwidth]{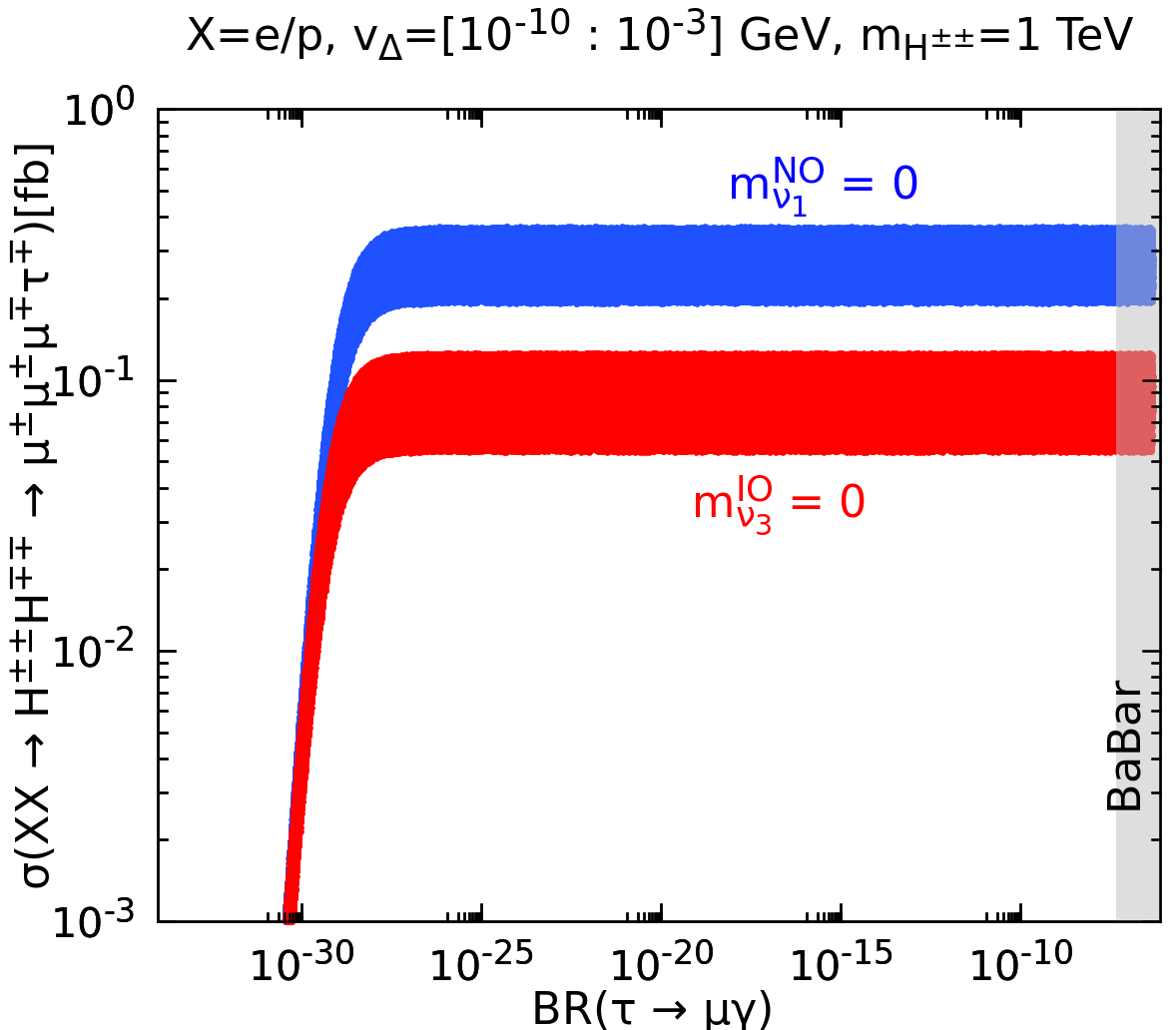}
\end{center}
\caption{
  Complementary cLFV probes: four-lepton cross-sections versus rare decay branching fractions $\text{BR}(\mu\to e\gamma)$ and $\text{BR}(\tau\to\mu\gamma)$. 
  Red bands correspond to {\bf IO}, while blue denote {\bf NO}. The lightest neutrino mass is taken to be zero.
  Oscillation parameters are varied within their $3\sigma$ ranges~\cite{deSalas:2020pgw,10.5281/zenodo.4726908}.
  The doubly-charged Higgs mass is $m_{H^{\pm\pm}}=1$~TeV and the triplet VEV lies in the range $10^{-10}\,\text{GeV}\leq v_\Delta \leq 10^{-3}\,\text{GeV}$.
  The gray bands are excluded by the MEG~\cite{MEG:2013oxv}~(top panel) and BaBar limits~\cite{BaBar:2009hkt}~(bottom panel).}
    \label{fig:BRLFVandCross0}
  \end{figure}

  That \clfv may be first observed as a high-energy phenomenon can be seen neatly from Fig.~\ref{fig:BRLFVandCross0}.
  In the top panel we display $\sigma(e^+ e^-\to H^{\pm\pm}H^{\mp\mp}\to eee\mu)$ versus $\text{BR}(\mu\to e\gamma)$ both for {\bf NO} and {\bf IO} with lightest neutrino mass $m_{\nu_{1(3)}}=0$.
  One sees that the four-lepton signature cross-section can be sizeable, even when the low-energy $\text{BR}(\mu\to e\gamma)$ decay branching ratio lies well below detectability. 
As before, for small $v_\Delta$, the Yukawa coupling is large and $H^{\pm\pm}$ decays predominantly to dileptons, hence cLFV can be sizeable. 
In the top panel, one sees that the cross section to $eee\mu$ final states is larger in the case of {\bf IO}, compared to {\bf NO}
and the difference becomes larger with a smaller mass of the lightest neutrino. 

Similarly, in the bottom panel of Fig.~\ref{fig:BRLFVandCross0} we show the \clfvg 4-lepton cross section $\sigma(e^+ e^-\to H^{\pm\pm}H^{\mp\mp}\to \mu^\pm\mu^\pm \mu^\mp\tau^\mp)$
in terms of $\text{BR}(\tau\to \mu\gamma)$. 
We again sees that the flavour-violating four-lepton final state cross-section can be sizeable even for tiny values of $\text{BR}(\tau\to\mu\gamma)$. 

Notice that in Fig.~\ref{fig:BRLFVandCross} we have scanned over the full 3$\sigma$ range of the oscillation parameters
  and found that the cross sections distinguish the mass orderings for most neutrino mass values. Likewise, in Fig.~\ref{fig:BRLFVandCross0} the predicted ranges differ,
  except for tiny cLFV branching fractions.

\section{Summary and outlook}
\label{sec:outlook}

In the simplest seesaw mechanism one can probe neutrino oscillation physics at collider energies through the pattern of triplet Higgs boson decays.
These can probe not only the lightest neutrino mass and the ordering of the neutrino masses, but also the flavour structure
of the neutrino sector, paving the way to the reconstruction of neutrino oscillation parameters at collider experiments.

For example, Fig.~\ref{fig:Atmos} illustrates how the decay pattern of the triplet Higgs that mediates neutrino mass generation may probe the octant of the atmospheric mixing angle. 
This can be tested at a high energy hadron colliders such as the FCC, as well as future $e^+ e^-$ colliders such as ILC, CLIC or CEPC in China.

Likewise, Fig.~\ref{fig:BRLFVandCross} shows how the rates for four-lepton final-state events coming from pair-producing the doubly-charged Higgs 
may be used as a probe of the light neutrino mass ordering, illustrating how high-energy signatures clearly complement neutrino oscillation studies.

Last, but not least, Fig.~\ref{fig:BRLFVandCross0} clearly suggests that \clfv could be observed first as a high-energy phenomenon,
since the corresponding signal cross section can be sizeable even when low-energy rare processes, such as $\mu \to e\gamma$, have negligible rates.
In short, high-energy probes clearly complement low-energy searches for \clfv at high-intensity facilities.

The results found here illustrate the complementarity and interplay of the high-energy and high-intensity frontiers in particle physics,
providing encouragement for dedicated simulation studies to evaluate the potential of these proposed facilities in probing the neutrino sector.\\

\begin{acknowledgments}
  The work of J.V. is supported by the Spanish grants PID2020-113775GB-I00 (AEI/10.13039/501100011033) and PROMETEO/2018/165 (Generalitat Valenciana).
  O. G. M. and G. S. G. were supported by CONACYT-Mexico under grant A1-S-23238. O. G. M. has been supported by SNI (Sistema Nacional de Investigadores).
  S.M. has been supported by KIAS Individual Grants (PG086001) at Korea Institute for Advanced Study. X.J.X has been supported in part by the National Natural Science Foundation of China under grant No. 12141501.
\end{acknowledgments}

\bibliographystyle{utphys}
\providecommand{\href}[2]{#2}\begingroup\raggedright\endgroup

\end{document}